%% file: main.tex
\newcommand{\CLASSINPUTinnersidemargin}{18mm}
\newcommand{\CLASSINPUToutersidemargin}{12mm}
\newcommand{\CLASSINPUTtoptextmargin}{20mm}
\newcommand{\CLASSINPUTbottomtextmargin}{25mm}
\documentclass[10pt,conference,a4paper]{IEEEtran}

\input{packages}

\input{macros}

\begin{document}

\title{Analysis and Improvements of the Sender Keys\\Protocol for Group Messaging}

\author{\IEEEauthorblockN{\textbf{David Balbás}}
\IEEEauthorblockA{IMDEA Software Institute,\\Universidad Politécnica de Madrid\\
 Spain\\
\url{david.balbas@imdea.org}}
\and
\IEEEauthorblockN{\textbf{Daniel Collins}}
\IEEEauthorblockA{EPFL\\
 Switzerland\\
\url{daniel.collins@epfl.ch}}
\and
\IEEEauthorblockN{\textbf{Phillip Gajland}}
\IEEEauthorblockA{Max Planck Institute for Security \& Privacy,\\
Ruhr-University Bochum\\
Germany\\
\url{phillip.gajland@mpi-sp.org}}}

\maketitle

\input{abstract}


\input{introduction}
\input{contents}

\section*{Acknowledgements}

This work has received funding from the European Research Council (ERC) under the European Union’s Horizon 2020 research and innovation program under project PICOCRYPT (grant agreement No. 101001283), and by a grant from Nomadic Labs and the Tezos foundation. The last author was supported by DFG under Germany’s Excellence Strategy - EXC 2092 CASA - 390781972. We are grateful to the reviewers of RECSI 2022 for their helpful comments on earlier versions of this work.

\bibliographystyle{TransactionsBibliography/IEEEtran}
\bibliography{abbrev0, crypto, extra}


\end{document}

%% file: packages.tex
\usepackage[T2A,T1]{fontenc}

\usepackage[dvipsnames]{xcolor}
\usepackage[colorlinks,backref=page,citecolor=OliveGreen,linkcolor=BrickRed]{hyperref}

\usepackage[utf8]{inputenc}

\usepackage[british]{babel}
\usepackage{multicol}
\usepackage[linewidth=.8pt]{mdframed}
\usepackage{caption}

\usepackage{amsmath, amsthm, mathtools, amsfonts, amssymb}
\usepackage{bm}

\usepackage{algorithm}
\usepackage{algorithmicx}
\usepackage[noend]{algpseudocode}

\expandafter\def\csname ver@etex.sty\endcsname{}
\usepackage[lambda,advantage,operators,sets,adversary,landau,probability,notions,logic,ff,mm,primitives,events,complexity,asymptotics,keys]{cryptocode}
\createprocedurecommand{procedure}{}{}{linenumbering,codesize=\scriptsize}

\usepackage{tikz}
\usepackage{tikz-cd}
\usetikzlibrary{arrows.meta}
\usetikzlibrary{automata,positioning}
\usetikzlibrary{graphs,graphs.standard,quotes}
\usetikzlibrary{shapes.multipart}

\usepackage{graphicx}
\usepackage[font=scriptsize]{caption}
\usepackage{subcaption}
\usepackage{array}
\usepackage{tabu}
\usepackage{wrapfig}
\usepackage{lipsum}
\usepackage{float}
\usepackage{multirow}
\usepackage{svg}
\usepackage{makecell}
\usepackage{microtype}

\usepackage{times}

\DeclareGraphicsExtensions{.png,.eps,.ps,.pdf}

\usepackage{url}

\usepackage{bm}

%% file: macros.tex
\newcommand{\G}{\bm{G}}
\newcommand{\gen}{\pcalgostyle{Gen}}

\newcommand{\GM}{\pcalgostyle{GM}}
\newcommand{\GMDef}{\GM\coloneqq(\init, \send, \recv, \exec, \proc)}

\newcommand{\ID}{\pcalgostyle{ID}}
\newcommand{\IDs}{\bm{IDs}}
\newcommand{\kdf}{\pcalgostyle{KDF}}
\newcommand{\ME}{\pcalgostyle{ME}}

\renewcommand{\state}{\gamma}
\newcommand{\m}{m}

\newcommand{\SenderK}{\pcalgostyle{send \pcmathhyphen k}}

\newcommand{\init}{\pcalgostyle{Init}}
\newcommand{\send}{\pcalgostyle{Send}}
\newcommand{\recv}{\pcalgostyle{Recv}}
\newcommand{\exec}{\pcalgostyle{Exec}}
\newcommand{\proc}{\pcalgostyle{Proc}}

\newcommand{\ct}{c}
\newcommand{\cmd}{\texttt{cmd}}
\newcommand{\add}{\texttt{add}}
\newcommand{\rem}{\texttt{rem}}
\newcommand{\crt}{\texttt{crt}}
\newcommand{\signseck}{\pcalgostyle{ssk}}
\newcommand{\ssk}{\signseck}
\newcommand{\spk}{\pcalgostyle{spk}}

\newcommand{\mk}{\pcalgostyle{mk}}
\newcommand{\ck}{\pcalgostyle{ck}}

\renewcommand{\O}{\mathcal{O}}
\newcommand{\pcassert}{\ensuremath{\mathbf{assert}}}

\renewcommand{\verify}{\pcalgostyle{Ver}}

\renewcommand{\sample}{
\mathrel{\vbox{\offinterlineskip\ialign{
			\hfil##\hfil\cr
			\hspace{0.1em}$\scriptscriptstyle\$$\cr
			$\leftarrow$\cr
		}}}
}

%% file: abstract.tex
\begin{abstract}
    Messaging between two parties and in the group setting has enjoyed widespread attention both in practice, and, more recently, from the cryptographic community.
    One of the main challenges in the area is constructing secure (end-to-end encrypted) and efficient messaging protocols for group conversations.
    The popular messaging applications WhatsApp and Signal utilise a protocol in which, instead of sharing a single group key, members have individual \textit{sender keys}, which are shared with all other group members.
    The Sender Keys protocol is claimed to offer forward security guarantees.
    However, despite its broad adoption in practice, it has never been studied formally in the cryptographic literature.

    In this paper we present the first analysis of the Sender Keys protocol along with some prospective improvements.
    To this end, we introduce a new cryptographic primitive, develop a game-based security model, present a security analysis in the passive and active settings, and propose several improvements to the protocol.
\end{abstract}

%% file: introduction.tex
\section{Introduction}\label{sec:introduction}

Messaging applications such as WhatsApp, Facebook Messenger, Signal and Telegram have enjoyed widespread adoption and form an integral part of communications for billions of people.
All of the aforementioned applications rely, to a varying degree, on cryptography to provide diverse forms of authenticity and secrecy.

Among end-to-end encrypted messaging solutions (this excludes Telegram and Facebook Messenger by default, among others), there exist diverse cryptographic solutions.
For two-party messaging, Signal's Double Ratchet Protocol~\cite{doubleratchet} is the most popular choice in practice, and many solutions also exist in the cryptographic literature~\cite{C:JaeSte18, EPRINT:PoeRos18, IWSEC:DurVau19, EC:AlwCorDod19, AC:BalRosVau20}.
For group messaging, the naive solution, as used by Signal Messenger for small groups, of adopting Double Ratchet sessions among every pair of group members does not scale well.
Thus, recent work such as the Messaging Layer Security (MLS) standardization effort~\cite{ietf-mls-protocol-14} aims to construct secure group messaging protocols where the complexity of group operations (adding and removing members, updating key material) is sublinear in the group size~\cite{Bhargavan2018, C:ACDT20, SP:KPWKCCMYAP21, TCC:ACJM20, CCS:ACDT21}.

Nevertheless, the popular messaging applications WhatsApp and Signal (for large groups) use a protocol for group messaging~\cite{whatsapp, signalrepo} that does not involve sharing a unique group key that evolves over time.
This differs from MLS, and from the group key agreement abstraction followed there~\cite{C:ACDT20, SP:KPWKCCMYAP21, TCC:ACJM20}.
This protocol, called Sender Keys, has not been formally studied in the literature despite its widespread adoption.

\subsection{Secure Group Messaging}\label{subsec:secure-group-messaging}

Two standard security notions prevail in the literature both for two-party and group messaging.
The first is \emph{forward security}~(FS), which protects the confidentiality of past messages in the event of a key exposure and can be achieved using just symmetric cryptography (for example by iteratively hashing symmetric keys).
The second is \emph{post-compromise security}~(PCS), which ensures that security can be restored after a key exposure in certain adversarial settings~\cite{CSF:CohCreGar16}, typically when the adversary is passive for some period of time. FS- and PCS-oriented key evolution mechanisms are commonly known as \emph{ratcheting}.

Both properties apply to the confidentiality and authenticity of sent messages and can be captured formally in a security game.
There exist different formalisations of security in the literature, but most of them model an adversarial Delivery Service (DS), the entity responsible for delivering messages between participants via the communication channel.
The adversary (modelling the DS) can act as an eavesdropper (with extended yet limited capabilities) as in~\cite{C:ACDT20}, as a semi-active adversary which can schedule messages arbitrarily~\cite{SP:KPWKCCMYAP21}, or as an active adversary that can inject messages~\cite{TCC:ACJM20, admins}.
In many protocols, including Sender Keys and MLS, the DS relies mainly on some centralized infrastructure (the \emph{central server} hereafter).

Some messaging protocols also require additional infrastructure to deal with user authentication or security.
This may include Public Key Infrastructure (PKI), or, in the case of Sender Keys, secure two-party messaging channels established between each pair of users.
Achieving security in multiple groups simultaneously is outside the scope of this work, and requires additional precautions detailed in~\cite{USENIX:CreHalKoh21}.

\subsection{Sender Keys}\label{subsec:sender-keys}

In a Sender Keys group $\G$, every user $\ID\in\G$ owns a so-called \emph{sender key} which is shared with all group members.
A sender key is a tuple $\SenderK = (\spk, \ck)$, where $\spk$ is a public signature key (with a private counterpart $\ssk$), and $\ck$ is a symmetric \emph{chain key}.
Every time a user $\ID$ sends a message $\m$ to the group, $\ID$ encrypts $\m$ using a \emph{message key} $\mk$ that is deterministically derived from its chain key $\ck$.
Upon message reception, group members also derive $\mk$ to decrypt the message.
Messages are authenticated by appending the sender's signature.

Forward security is provided by using a fresh message key for every message; every time a message is sent, the chain key is hashed forward using a key derivation function. In other words, chain keys are symmetrically ratcheted.

The protocol also requires that there exist confidential and authenticated two-party communication channels between every pair of users.
These are used for sharing sender keys in the event of parties being added or removed from the group.

\subsection{Contributions}\label{subsec:contributions}

The main scientific contributions of our paper are the following:

\begin{itemize}
    \item We introduce a new cryptographic primitive, Group Messenger ($\GM$), which is suitable for messaging protocols like Sender Keys that are not necessarily based on group key agreement.
    \item We formally describe Sender Keys, based on a code analysis of Signal's source code~\cite{signalrepo}, and WhatsApp's security white paper~\cite{whatsapp}.
    \item We present a security model for (single-group) Group Messenger.
    We do so via a security game that considers an active adversary who can interact with several oracles.
    The game is parametrised by a cleanness predicate that captures forward-secure group messaging.
    \item We carry out a security analysis for passive and active adversaries and detail prospective fixes and improvements to the Sender Keys protocol.
\end{itemize}

We note that this is a preliminary and shortened version of our work.

%% file: contents.tex
\section{Primitive Syntax}\label{sec:primitive-syntax}

Unless otherwise stated, all algorithms are probabilistic, and $(x_1,\dots)\sample\adv(y_1,\dots)$ is used to denote that $\adv$ returns $(x_1,\dots)$ when run on input $(y_1,\dots)$.
Blank values are represented by $\bot$.
We denote the security parameter by $\secpar$ and its unary representation by $\secparam$.
We also define the state $\state$ of a user $\ID$ as the data required by $\ID$ for protocol execution, including message records, group-related variables, and cryptographic material.

We introduce a cryptographic primitive that we call \emph{Group Messenger} $\GMDef$, similar to other group messaging abstractions such as Continuous Group Key Agreement~(CGKA)~\cite{C:ACDT20}.
In contrast to CGKA, our primitive does not model key agreement (as this does not neatly capture the Sender Keys protocol), but rather sending and receiving messages.
The syntax is as follows.
\begin{itemize}
    \item $\state\sample\init(\secparam,\ID)$: Given the security parameter $\secparam$ and a user identity $\ID$, the probabilistic initialisation algorithm returns an initial state $\state$.
    \item $(C, \state') \sample \send(\m, \state)$: Given a message $\m$, and a state $\state$, the probabilistic sending algorithm returns a ciphertext $C$ and a new state $\state'$.
    \item $(\m, \state') \gets \recv(C, \state)$: Given a ciphertext $C$, and a state $\state$, the deterministic receiving algorithm returns a message $\m$ and a new state $\state'$.
    \item $(C, \state') \sample \exec(\cmd, \IDs, \state)$: Given a command $\cmd \in \{\crt, \add, \rem\}$, a list of user identities $\IDs$, and a state $\state$, the probabilistic execution algorithm returns a ciphertext $C$ and a new state $\state'$.
    \item $\state' \gets \proc(C, \state)$: Given a ciphertext $C$, and a state $\state$, the deterministic processing algorithm returns a new state $\state'$.
\end{itemize}

Note that there are separate algorithms for sending / receiving application messages and for executing / processing changes to the group.
Furthermore, should two-party protocols be required under the hood to implement the group primitive (this is not the case for CGKAs such as TreeKEM~\cite{C:ACDT20}), then these are outside the scope of this definition.

\section{Protocol Description}\label{sec:protocol-description}

In this section, we introduce a formal description of the Sender Keys protocol in accordance with our Group Messenger syntax.

\subsection{Protocol Setup}\label{subsec:protocol-setup}

\subsubsection{Central server}
The Delivery Service (DS) relies on a central server which provides total ordering of messages and authenticates users initially (i.e. it acts as a PKI).
In practice, the DS is also responsible for managing two-party channels.

\subsubsection{Two-party channels}
The protocol assumes that there exist authenticated and secure two-party communication channels (for example using Signal's Double Ratchet protocol~\cite{doubleratchet} as done in WhatsApp~\cite{whatsapp}) between every pair of protocol users.
This assumption can be realized via the Delivery Service and asynchronous, PKI-aided key-exchange mechanisms such as X3DH~\cite{x3dh}.

\subsubsection{Primitives}
The protocol uses standardised~\cite{KraBelCan97,RFC5869} underlying cryptographic primitives:
\begin{itemize}
    \item Two different Key Derivation Functions~(KDF) $H_1,H_2:\bin^\secpar\to\bin^\secpar$.
    These are used to derive message keys $\mk$ and chain keys $\ck$, respectively.
    \item A symmetric encryption scheme $(\enc, \dec)$.
    \item A digital signature scheme $(\gen, \sig, \verify)$.
\end{itemize}

In Signal's implementation of Sender Keys~\cite{signalrepo}, the KDFs are instantiated as $H_1(m) \coloneqq \mathsf{HMAC}(\texttt{0x01}, m)$ and $H_2(m) \coloneqq \mathsf{HMAC}(\texttt{0x02}, m)$.
We note that in Signal's two-party sessions signatures are not used.
Instead, messages are authenticated by computing a MAC based on a function of the message key.

In WhatsApp~\cite{whatsapp}, the $\mathsf{HMAC}$ used for the KDF is \texttt{HMAC-SHA256}, the symmetric encryption scheme is \texttt{AES-256} in CBC mode, and the signature scheme is \texttt{ECDSA} with \texttt{Curve25519}.

\subsection{State}\label{subsec:state}
The state $\state$ of a user $\ID$ contains a sender key $\SenderK\coloneqq(\spk,\ck)$, where $\spk$ denotes the the signature public key and $\ck$ the chain key, as well as a secret signature key $\ssk$, each belonging to $\ID$.
The state also maintains a list of current group members $\G$.
Additionally, for each user $\ID\in\G$ the sender key $\SenderK_{\ID}$, a list of skipped message keys $\{\mk_{\ID}^i,\dots\}$, used for out-of-order delivery, and the counter $i$ are stored.
If $\ID$ leaks its state, we say that it suffered a \emph{state compromise}.

\subsection{Algorithms}\label{sec:algorithms}
\begin{figure}[htb]
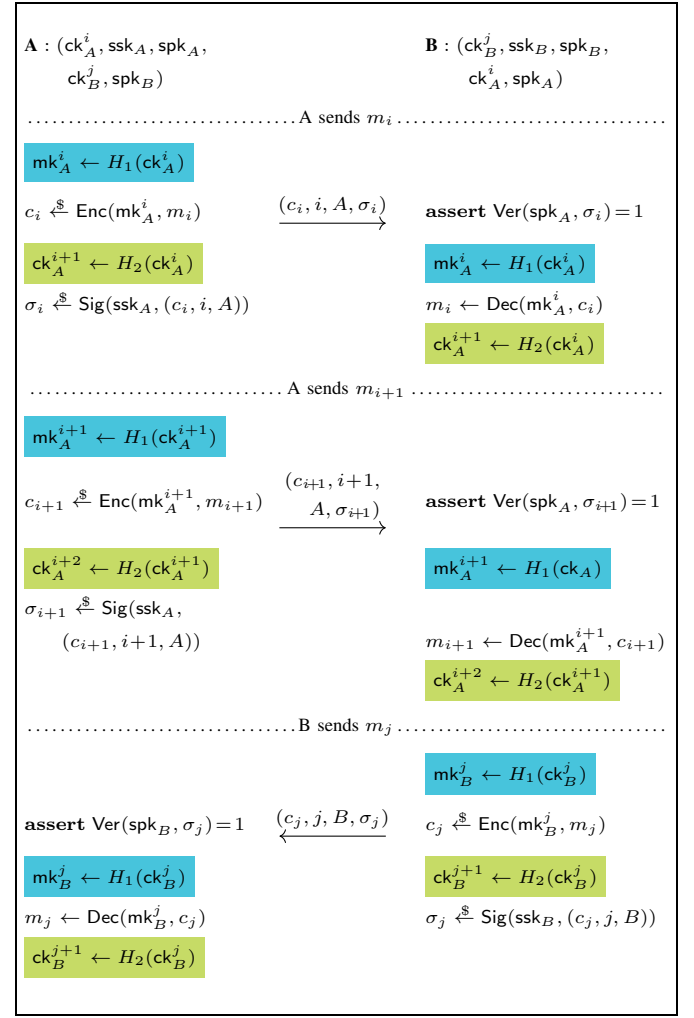

    \centering
    \begin{pcvstack}[boxed, space = 0.5em]
        \procedureblock[codesize=\scriptsize]{}{ \\
        \textbf{A}: (\ck^i_A,\ssk_A,\spk_A, \> \> \textbf{B}: (\ck^j_B,\ssk_B,\spk_B,\\
        \quad\quad\;\ck_B^j,\spk_B) \> \>\quad\quad\;\ck_A^i,\spk_A)\pclb
        \pcintertext[dotted]{A sends $m_i$}
        \colorbox{SkyBlue}{$\mk^i_{A} \gets H_1(\ck^i_{A})$} \> \> \\
        \ct_i \sample \enc(\mk^i_{A}, \m_i) \> \sendmessageright*[1.4cm]{(\ct_i, i,A,\sigma_i)} \>  \pcassert\;\verify(\spk_A,\sigma_i)\!=\!1 \\
        \colorbox{SpringGreen}{$\ck^{i+1}_A \gets H_2(\ck^i_A)$} \> \> \colorbox{SkyBlue}{$\mk^i_{A} \gets H_1(\ck^i_{A})$} \\
        \sigma_i \sample \sig(\ssk_A, (\ct_i, i, A)) \> \> \m_i \gets \dec(\mk^i_{A}, \ct_i) \\
        \> \> \colorbox{SpringGreen}{$\ck^{i+1}_A \gets H_2(\ck^{i}_A)$} \pclb
        \pcintertext[dotted]{A sends $\m_{i+1}$}
        \colorbox{SkyBlue}{$\mk^{i+1}_{A} \gets H_1(\ck^{i+1}_{A})$} \> \> \\
        \ct_{i+1} \sample \enc(\mk^{i+1}_{A}, \m_{i+1}) \> \sendmessageright*[1.4cm]{(\ct_{i\!+\!1},i\!+\!1,\\A,\sigma_{i\!+\!1})} \> \pcassert\;\verify(\spk_A,\sigma_{i\!+\!1})\!=\!1 \\
        \colorbox{SpringGreen}{$\ck^{i+2}_A \gets H_2(\ck^{i+1}_A)$} \> \>  \colorbox{SkyBlue}{$\mk^{i+1}_{A} \gets H_1(\ck_{A})$} \\
        \sigma_{i+1} \sample \sig(\ssk_A,\\ \quad\quad(\ct_{i+1},i\!+\!1, A)) \> \> \m_{i+1} \gets \dec(\mk^{i+1}_{A}, \ct_{i+1}) \\
        \> \> \colorbox{SpringGreen}{$\ck^{i+2}_A \gets H_2(\ck^{i+1}_A)$} \pclb
        \pcintertext[dotted]{B sends $\m_j$}
        \> \> \colorbox{SkyBlue}{$\mk^{j}_{B} \gets H_1(\ck^j_{B})$} \\
        \pcassert\;\verify(\spk_B,\sigma_j)\!=\!1  \> \sendmessageleft*[1.4cm]{(\ct_{j}, j,B,\sigma_j)} \> \ct_{j} \sample \enc(\mk^{j}_{B}, \m_{j}) \\
        \colorbox{SkyBlue}{$\mk^{j}_{B} \gets H_1(\ck^j_{B})$} \> \>  \colorbox{SpringGreen}{$\ck^{j+1}_B \gets H_2(\ck^j_B)$} \\
        \m_{j} \gets \dec(\mk^{j}_{B}, \ct_{j}) \> \>  \sigma_j \sample \sig(\ssk_B, (\ct_j, j, B))\\
        \colorbox{SpringGreen}{$\ck^{j+1}_B \gets H_2(\ck^j_B)$} \> \>  \\
        }
    \end{pcvstack}
    \caption{Sending/receiving messages between two group members for a three-message (in-order) conversation.
    Ephemeral message keys $\mk$ are deleted immediately after use.
    A's initial sender key is $(\ck^i_A, \ssk_A)$ and B's initial sender key is $(\ck^j_B, \ssk_B)$.}
    \label{fig:example}
\end{figure}

We describe the Sender Keys protocol according to our Group Messenger primitive defined in Section~\ref{sec:primitive-syntax}.
The description follows Signal's reference implementation~\cite{signalrepo} regarding sender key ratcheting as well as message encryption and decryption.
The details of the $\exec$ and $\recv$ algorithms are inferred from~\cite{whatsapp}, but we cannot assert that our interpretation is entirely faithful to WhatsApp's implementation.
A simplified example of a 3-message conversation is shown in Figure~\ref{fig:example}.

\subsubsection{State initialization}
The $\init$ algorithm initializes the state variables of users; in practice this is done at install time.

\subsubsection{Group creation}
The creation of a group is carried out via the $\exec(\crt, \G, \state)$ algorithm which takes a list of prospective members $\G\coloneqq\{\ID_1,\ldots, \ID_{|\G|}\}$ as input.
All parties are assumed to have pre-established two-party communication sessions.
The group creator $\state.\ME$ generates a chain key $\ck\sample\bin^\secpar$ and a signature key pair $(\state.\spk, \state.\ssk) \sample \gen(\secparam)$.
Then, it sends its sender key $\state.\SenderK=(\state.\spk, \ck)$ to each $\ID\in\G$ individually via their secure two-party channel.

\subsubsection{Message sending}
To send the $i$\textsuperscript{th} message $m_i$, a user $\state.\ME$ calls the $\send(m_i,\state)$ algorithm which does the following:
\begin{itemize}
    \item Derive a new message key $\mk$ from the symmetric part of its sender key (i.e. the chain key) as $\mk^i_\ME \gets H_1(\ck^i_\ME)$.
    \item Encrypt the message $m_i$ as $\ct_i \sample \enc(\mk^i_\ME, m_i)$.
    \item Ratchet the chain key $\ck$\footnote{Note that this practice is safer (better FS) than first evolving the chain key and then deriving a message key, since it allows for the immediate deletion of the chain key used to derive the message key.} as $\ck_\ME^{i+1}\gets H_2(\ck^i_\ME)$.
    \item Jointly sign the ciphertext $\ct_i$, the message index $i$ and the sender's identity $\ME$ as $\sigma_i \sample \sig(\ssk, (\ct_i, i, \ME))$
    \item Send $C\coloneqq(\ct_i, i, \ME, \sigma_i)$ to all group members via the DS.
\end{itemize}
If $\state.\SenderK = \bot$, then $\state.\ME$ must first generate a fresh sender key and distribute it as in the group creation.
This occurs every time a member sends their first message after entering the group or after a member removal.
We emphasise that ciphertexts are not sent over the preexisting two-party channels that are used to communicate sender keys, but rather over the network (i.e., via the Delivery Service) itself.

\subsubsection{Message reception}

Upon reception of a message $C=(\ct_i, i, \ID, \sigma_i)$, the receiver calls $\recv(C, \state)$. First, $\recv$ verifies $\ID$ and the signature as $\verify(\SenderK[\ID].\spk, \sigma_i, (\ct_i, i, \ID))$ (note that if $\SenderK[\ID]=\bot$, the receiver must wait until a new sender key is sent by $\ID$). If the check passes, then the algorithm proceeds as follows:
\begin{itemize}
    \item If $\SenderK[\ID].\ck$ is at iteration $i$:

    derive $\mk_i \gets H_1(\SenderK[\ID.\ck])$,

    decrypt $\ct_i$ as $m_i\gets \dec(\mk_i, \ct_i)$ and erase $\mk_i$,

    and refresh $\SenderK[\ID.\ck] \gets H_2(\SenderK[\ID.\ck])$.
    \item If $\SenderK[\ID].\ck$ is at iteration $j<i$, refresh the chain key $i-j$ times as $\SenderK[\ID.\ck] \gets H_2(\SenderK[\ID.\ck])$ while storing message keys $\mk_j, \ldots \mk_{i-1}$ (up to $N_{\mathsf{max}}$ keys). Then, obtain $\mk_{i}$, decrypt $\ct_i$, and erase $\mk_i$.
    \item If $\SenderK[\ID].\ck$ is at iteration $j>i$, search for a stored $\mk_j$, attempt to decrypt $\ct_i$, and erase $\mk_j$. If unsuccessful, output $\bot$.
\end{itemize}

\subsubsection{Membership changes}
To add a new user $\ID$ to the group, a member calls $\exec(\texttt{add}, \{\ID\}, \state)$.
This produces a notification message $T$ sent to all group members including $\ID$.
As mentioned earlier, $\ID$ only generates and sends its own sender key when they send their first message. $\ID$ will receive the sender key of every other member $\ID'$ when they speak in the group for the first time after $\ID$ joins. This is sent over the two-party channel of $\ID$ and $\ID'$.

To remove a user $\ID$ from the group, a member calls $\exec(\texttt{rem}, \{\ID\}, \state)$ and sends the notification $T$ to all members.

\subsubsection{Message processing}
Group changes are processed via $\proc(T, \state)$. If a user $\ID$ is added, $\state.\ME$ simply updates the list of group members $\state.G$, and sends $\ID$ his own sender key via the two-party channel when he speaks again in the group.
If $\ID$ is removed, $\state.\ME$ deletes all sender keys, including his own.
In this scenario, the group ``starts over", namely all members must generate and send a new sender key (i.e. a new signature key pair and symmetric key) to the members.

\section{Security Model}\label{sec:security-model}
We propose a model of security for our Group Messenger primitive that captures the security suitable for an authenticated and forward-secure group messaging scheme.
We introduce a game played between a probabilistic polynomial time ($\ppt$) adversary $\adv$ and a challenger.

\subsection{Game Description}\label{sec:game-description}
At the beginning of the game, a bit $b$ is uniformly sampled which parametrises the game.
To win, the adversary either has to guess $b$ or carry out a successful forgery in a \emph{clean} protocol run.
The game is parameterised by a \emph{cleanness} predicate (sometimes safety~\cite{C:ACDT20} predicate) that captures the exact security of the protocol, namely the authenticity and confidentiality of group messages.

In this work, we assume that the two-party communication channels used for sending sender keys are perfectly secure; i.e., always confidential and authenticated.
This assumption is not easily met in practice, but allows us to capture the essence of Sender Keys alone.

For simplicity, the game starts with a pre-established group $\G$ where every member $\ID\in\G$ already has everyone's honestly generated sender key.
Then, $\adv$ can interact with several oracles:
\begin{itemize}
    \item $\O^{\mathsf{Challenge}}(\ID, \m_0, \m_1)$. The adversary receives a ciphertext $C_b$ corresponding to $\ID$ sending $\m_b$, i.e., the output of $C_b\gets \send(\m_b, \state_\ID)$.
    \item $\O^{\mathsf{Send}}(\ID, \m)$. $\ID$ sends an application-level message $\m$ using the $\send$ algorithm, producing a ciphertext $C$.
    \item $\O^{\mathsf{Receive}}(\ID, \ID', C)$. $\ID$ calls $\recv$ on an application-level ciphertext $C$ claimed to be from user $\ID'$.
    If $C$ has not been generated honestly via the $\O^{\mathsf{Send}}(\ID, \m)$ oracle and receiving is successful, $b$ is leaked to $\adv$.
    \item $\O^{\mathsf{Add}}(\ID, \ID')$. $\ID$ adds $\ID'$ to the group via $\exec$, generating a control message $T$.
    \item $\O^{\mathsf{Remove}}(\ID, \ID')$. $\ID$ removes $\ID'$ from the group via $\exec$, generating a control message $T$.
    \item $\O^{\mathsf{Deliver}}(\ID, T)$. A control message $T$ is delivered to $\ID$ who calls $\proc$.
    \item $\O^{\mathsf{Expose}}(\ID)$. The current state $\state$ of $\ID$ leaks to $\adv$.
    \item $\O^{\mathsf{ExpMK}}(\ID, i)$. The $i$-th message key $\mk$ of $\ID$ leaks to $\adv$. No message encrypted under this key can be challenged (neither before nor after exposure).
\end{itemize}

After $q$ oracle queries, the adversary outputs a guess $b'$ of $b$.
Note that $\adv$ can win the game either by guessing a challenge correctly or by injecting a forged message via $\O^{\mathsf{Receive}}$ successfully.

\subsection{Cleanness} \label{sec:cleanness}
For the particular case of Sender Keys, we describe the cleanness predicate which defines the following conditions for a valid game:
\begin{itemize}
    \item We define the event $\mathsf{refresh}(\ID)$ that occurs when: 1) some member has been removed from the group, and 2) member $\ID$ processes this change (note the lack of a PCS update option).
    \item After exposing \emph{any} user $\ID$, all adversarial calls to $\O^{\mathsf{Challenge}}$ on future messages are disallowed until $\mathsf{refresh}(\ID)$ occurs for \emph{every} $\ID \in \G$.
    This extends to all challenges on skipped messages (out-of-order) that $\ID$ has not received at exposure time.
    \item After exposing \emph{a specific} user $\ID'$, $\adv$ cannot win the game by impersonating $\ID'$ via $\O^{\mathsf{Receive}}(\ID, \ID', C)$ for a forgery $C$ until a new $\mathsf{refresh}(\ID)$ event occurs.
\end{itemize}
We remark that our security notion is adaptive insofar as users can adaptively expose users.
Under our cleanness predicate, we consider limited injection queries, i.e., partially active security.
Our modelling further assumes that the underlying two-party channels are perfectly secure, and thus we leave it as important future work to examine security where, e.g., state exposures on the underlying channels are allowed and the consequent security guarantees are captured.

\section{Security}
We claim that Sender Keys, as described in Section~\ref{sec:protocol-description}, is secure with respect to our security model.
However, the security captured by our cleanness predicate is sub-optimal, in the sense that forward security can be strengthened for authentication, as we introduce in Section \ref{sec:activeattacks}.
Here we introduce a security analysis, but leave a security proof and more accurate modelling for future work.

\subsection{Passive adversaries}
For a (semi-)passive adversary which does not attempt to inject messages via $\O^{\mathsf{Receive}}$ (but can still schedule messages arbitrarily), we claim that Sender Keys is secure with respect to the cleanness predicate, given that the symmetric encryption scheme is IND-CPA secure.
Towards proving this:
\begin{itemize}
    \item If the $\kdf$ is a one-way function, message keys $\mk_i$ can be exposed independently; the compromise of a message key never affects the confidentiality of other keys or messages. Hence, giving adversarial access to $\O^{\mathsf{ExpMK}}(\ID, i)$ does not impact the cleanness predicate.
    \item Also assuming one-wayness of the $\kdf$, forward secrecy holds trivially except for out-of-order messages.
    \item Assuming that the two-party channels are secure, all users recover from state exposure via $\O^{\mathsf{Expose}}(\ID)$ after a removal is made effective. Note that, outside our model, security of the two-party channels may also degrade after a state exposure, leaving room for further attacks.
\end{itemize}

\subsection{Single- vs multi-key}
In a Sender Keys group, each user is associated with a different symmetric key and thus the state comprises $O(n)$ secret material at all times.
Since users encrypt and then hash forward using their own key when sending each message, users can safely send messages concurrently and with some inter-member message reordering.

For large groups, however, this scaling behaviour may represent a bottleneck.
Consequently, one can envision trade-offs between the amount of concurrency supported and the amount of secret material required to be stored at a given point a time.
The other extreme of the spectrum would be when all users maintain the \emph{same} single symmetric chain.
In situations where users are not expected to concurrently send messages this allows the secret state size to reduce to $O(1)$ without degrading security.
To deal with concurrency in this setting, a central server which rejects all but the first (for example) message and requests re-transmission for other users could then be employed.

In MLS, a new group secret (chosen by a single user) is established each epoch, from which point all $O(n)$ application keys are derived for a given point in time; MLS additionally supports out-of-order message delivery within a given epoch.
In Sender Keys, each user chooses their own key; thus, the security of ciphertexts in the presence of a passive adversary is contingent upon users initially sampling their key with enough entropy.

\subsection{Active attacks} \label{sec:activeattacks}

In an active adversary scenario, the security of Sender Keys is sub-optimal. In particular, we note two issues. First, consider a simple group $\G=\{\ID_1, \ID_2\}$ and the following sequence of oracle queries:
\begin{itemize}
    \item $q_1= \O^{\mathsf{Send}}(\ID_1, \m)$ which generates the $i$-th message $C$ encrypted under $\mk$ and signed under $\ssk_{1}$.
    \item $q_2= \O^{\mathsf{Expose}}(\ID_1)$, where $\adv$ obtains $\ssk_{1}$, but not $\mk$.
    \item $q_3= \O^{\mathsf{ExpMK}}(\ID_1, i)$, leaking $\mk$.
    \item $\adv$ crafts $\ct' = Enc(\m', \mk)$, signs it under $\ssk_{1}$ and forges a $C'$.
    \item $q_4= \O^{\mathsf{Deliver}}(\ID_2, C')$, as the $i$-th message.
\end{itemize}
Note that $q_4$ is a forbidden query by our cleanness predicate in Section \ref{sec:cleanness}. $q_4$ attempts to inject a message that corresponds to key material utilized \emph{before} the state exposure, hence one can envision stronger forward security where queries like $q_4$ are allowed.
In this case, the Sender Keys adversary would win the game.
We describe how to achieve such stronger security in Section \ref{sec:modifications} by strengthening signatures.

Technically, the query $q_3$ can be replaced by $\O^{\mathsf{Expose}}(\ID_k)$, or even be omitted as the adversary can still create a valid forgery by altering the metadata in $C$ (such as the sender's identity) without crafting a new $\ct'$.
This attack can occur naturally if $\ID_2$ is offline when $\m$ is first sent.

An attack of a similar nature can also occur if the same signature key is re-used across groups, and they are refreshed at different times, as pointed out in~\cite{USENIX:CreHalKoh21}.

The second issue we note is that the implementation of the $\exec$ algorithm can be problematic if messages are not authenticated correctly.
This led to attacks in the past such as the \emph{burgle into the group} or the \emph{acknowledgement forgery} attacks in~\cite{rosler2018more}.
Securing control messages and group membership changes is possible as introduced in~\cite{admins}.

\subsection{Proposed Modifications and Tweaks} \label{sec:modifications}

\subsubsection{Ratcheting signature keys} The attack shown in the previous section can be mitigated if signature keys are also ratcheted. A simple fix is to introduce a \emph{chain} of signature keys, where an ephemeral signature key is created every time a message or block of messages is sent.

Let $(\ssk, \spk)$ be $\ID$'s signature key pair, where $\spk$ is part of its Sender Key. Then, before sending a new message $m$ to the group, $\ID$ can generate a new key pair $(\ssk', \spk')\sample \gen(\secparam)$. Then, $\ID$ can do as follows:
\begin{itemize}
    \item Encrypt the $i$\textsuperscript{th} message $m$ with the corresponding $\mk$, $\ct \sample \enc(\mk, m)$.
    \item Sign $C \gets (c, i, \ID, \spk')$ as $\sigma \sample \sig(\ssk, C)$
    \item Send the tuple $(\sigma, C)$ to the group.
    \item Replace $\ssk$ by $\ssk'$.
\end{itemize}
Upon reception of a message, $\ID'$ will:
\begin{itemize}
    \item Verify the signature as $\verify(\spk, \sigma, C)$ and decrypt the ciphertext $c$.
    \item Replace $\spk$ by $\spk'$ in $\ID$'s sender key.
\end{itemize}

Note that this countermeasure involves a notable overhead and entropy consumption (although only for the sender), so it may not be desirable in all scenarios, or for all sent messages.
Users could also replace their signature keys on-demand or on a time schedule to trade more post-compromise security for performance.



\subsubsection{Randomness manipulation} We note that Sender Keys, as described in Section~\ref{sec:protocol-description}, is susceptible to randomness exposure and randomness manipulation attacks.
Namely, the adversary does not need to leak a member's state, but simply control the randomness used by the device, inhibiting any form of PCS.
Protection against this family of attacks can be attained at small cost if freshly generated keys are hashed with the state or with part of the state, as in~\cite{C:JaeSte18} and the classic NAXOS trick for authenticated key exchange~\cite{PROVSEC:LaMLauMit07}.

\subsubsection{Refresh option for PCS}\label{sec:refresh-option-for-pcs}
Post compromise security (PCS) is generally achieved when introducing fresh randomness by establishing new group secrets.
In the case of Sender Keys, it could be beneficial to establish new sender keys across the group at some intervals, for instance via an \emph{update} group operation, as noted in \cite{USENIX:CreHalKoh21}.
However, if only Alice updates her sender key, a passive adversary would still be able to eavesdrop on messages sent by any other group member.
Furthermore, as the sender key is group-specific, any messages sent by Alice in another group are also susceptible to eavesdropping.
We refer the reader to Appendix A of~\cite{USENIX:CreHalKoh21} for a more detailed discussion and conclude that the sender keys is not a suitable approach towards achieving a PCS-secure group messenger.

\section{Conclusion and Future Work}\label{sec:conclusion}

The Sender Keys protocol is a convenient protocol for group chats which is simple to implement, achieves a fair degree of end-to-end forward security, and deals well with concurrency.
Nevertheless, we remark that no security is gained by employing multiple sender keys with respect to a single group key given randomness used to generate the keys is honest (i.e., in our model).
Furthermore, forward security is sub-optimal for message authentication, which can be fixed by ratcheting signature keys.

Our results are under the assumption that two-party channels are secure, which is clearly not the case in practice.
Therefore, more powerful attacks may arise under a more realistic model where exposure of two-party channels is possible.
We also note that, in the real world, the security of messaging apps can also be broken in different ways than attacking the protocol.
Additional features such as conversation transcript backups and multi-device support highly increase the attack surface and should be avoided in applications where security is critical.

Our analysis leaves multiple directions for future work, such as a rigorous formalization of the security model with a fine-grained analysis of the two-party channels and user compromise, more precise cleanness predicates, and especially concrete security reductions.

%% file: main.bbl
\begin{thebibliography}{10}
\providecommand{\url}[1]{#1}
\csname url@samestyle\endcsname
\providecommand{\newblock}{\relax}
\providecommand{\bibinfo}[2]{#2}
\providecommand{\BIBentrySTDinterwordspacing}{\spaceskip=0pt\relax}
\providecommand{\BIBentryALTinterwordstretchfactor}{4}
\providecommand{\BIBentryALTinterwordspacing}{\spaceskip=\fontdimen2\font plus
\BIBentryALTinterwordstretchfactor\fontdimen3\font minus
  \fontdimen4\font\relax}
\providecommand{\BIBforeignlanguage}[2]{{%
\expandafter\ifx\csname l@#1\endcsname\relax
\typeout{** WARNING: IEEEtran.bst: No hyphenation pattern has been}%
\typeout{** loaded for the language `#1'. Using the pattern for}%
\typeout{** the default language instead.}%
\else
\language=\csname l@#1\endcsname
\fi
#2}}
\providecommand{\BIBdecl}{\relax}
\BIBdecl

\bibitem{doubleratchet}
\BIBentryALTinterwordspacing
M.~Marlinspike and T.~Perrin, ``The double ratchet algorithm,'' 2016. [Online].
  Available:
  \url{https://signal.org/docs/specifications/doubleratchet/doubleratchet.pdf}
\BIBentrySTDinterwordspacing

\bibitem{C:JaeSte18}
J.~Jaeger and I.~Stepanovs, ``Optimal channel security against fine-grained
  state compromise: The safety of messaging,'' in \emph{Advances in Cryptology
  -- {CRYPTO}~2018, Part~I}, ser. Lecture Notes in Computer Science, H.~Shacham
  and A.~Boldyreva, Eds., vol. 10991.\hskip 1em plus 0.5em minus 0.4em\relax
  Santa Barbara, CA, USA: Springer, Heidelberg, Germany, Aug.~19--23, 2018, pp.
  33--62.

\bibitem{EPRINT:PoeRos18}
B.~Poettering and P.~R{\"o}sler, ``Asynchronous ratcheted key exchange,''
  Cryptology ePrint Archive, Report 2018/296, 2018,
  \url{https://eprint.iacr.org/2018/296}.

\bibitem{IWSEC:DurVau19}
F.~B. Durak and S.~Vaudenay, ``Bidirectional asynchronous ratcheted key
  agreement with linear complexity,'' in \emph{IWSEC 19: 14th International
  Workshop on Security, Advances in Information and Computer Security}, ser.
  Lecture Notes in Computer Science, N.~Attrapadung and T.~Yagi, Eds., vol.
  11689.\hskip 1em plus 0.5em minus 0.4em\relax Tokyo, Japan: Springer,
  Heidelberg, Germany, Aug.~28--30, 2019, pp. 343--362.

\bibitem{EC:AlwCorDod19}
J.~Alwen, S.~Coretti, and Y.~Dodis, ``The double ratchet: Security notions,
  proofs, and modularization for the {Signal} protocol,'' in \emph{Advances in
  Cryptology -- {EUROCRYPT}~2019, Part~I}, ser. Lecture Notes in Computer
  Science, Y.~Ishai and V.~Rijmen, Eds., vol. 11476.\hskip 1em plus 0.5em minus
  0.4em\relax Darmstadt, Germany: Springer, Heidelberg, Germany, May~19--23,
  2019, pp. 129--158.

\bibitem{AC:BalRosVau20}
F.~Balli, P.~R{\"o}sler, and S.~Vaudenay, ``Determining the core primitive for
  optimally secure ratcheting,'' in \emph{Advances in Cryptology --
  {ASIACRYPT}~2020, Part~III}, ser. Lecture Notes in Computer Science,
  S.~Moriai and H.~Wang, Eds., vol. 12493.\hskip 1em plus 0.5em minus
  0.4em\relax Daejeon, South Korea: Springer, Heidelberg, Germany, Dec.~7--11,
  2020, pp. 621--650.

\bibitem{ietf-mls-protocol-14}
\BIBentryALTinterwordspacing
R.~Barnes, B.~Beurdouche, R.~Robert, J.~Millican, E.~Omara, and K.~Cohn-Gordon,
  ``{The Messaging Layer Security (MLS) Protocol},'' Internet Engineering Task
  Force, Internet-Draft draft-ietf-mls-protocol-14, May 2022, work in Progress.
  [Online]. Available:
  \url{https://datatracker.ietf.org/doc/html/draft-ietf-mls-protocol-14}
\BIBentrySTDinterwordspacing

\bibitem{Bhargavan2018}
\BIBentryALTinterwordspacing
K.~Bhargavan, R.~Barnes, and E.~Rescorla, ``{TreeKEM: Asynchronous
  Decentralized Key Management for Large Dynamic Groups A protocol proposal for
  Messaging Layer Security (MLS)},'' {Inria Paris}, Research Report, May 2018.
  [Online]. Available: \url{https://hal.inria.fr/hal-02425247}
\BIBentrySTDinterwordspacing

\bibitem{C:ACDT20}
J.~Alwen, S.~Coretti, Y.~Dodis, and Y.~Tselekounis, ``Security analysis and
  improvements for the {IETF} {MLS} standard for group messaging,'' in
  \emph{Advances in Cryptology -- {CRYPTO}~2020, Part~I}, ser. Lecture Notes in
  Computer Science, D.~Micciancio and T.~Ristenpart, Eds., vol. 12170.\hskip
  1em plus 0.5em minus 0.4em\relax Santa Barbara, CA, USA: Springer,
  Heidelberg, Germany, Aug.~17--21, 2020, pp. 248--277.

\bibitem{SP:KPWKCCMYAP21}
K.~Klein, G.~Pascual-Perez, M.~Walter, C.~Kamath, M.~Capretto, M.~Cueto,
  I.~Markov, M.~Yeo, J.~Alwen, and K.~Pietrzak, ``Keep the dirt: Tainted
  {TreeKEM}, adaptively and actively secure continuous group key agreement,''
  in \emph{2021 {IEEE} Symposium on Security and Privacy}.\hskip 1em plus 0.5em
  minus 0.4em\relax San Francisco, CA, USA: {IEEE} Computer Society Press,
  May~24--27, 2021, pp. 268--284.

\bibitem{TCC:ACJM20}
J.~Alwen, S.~Coretti, D.~Jost, and M.~Mularczyk, ``Continuous group key
  agreement with active security,'' in \emph{TCC~2020: 18th Theory of
  Cryptography Conference, Part~II}, ser. Lecture Notes in Computer Science,
  R.~Pass and K.~Pietrzak, Eds., vol. 12551.\hskip 1em plus 0.5em minus
  0.4em\relax Durham, NC, USA: Springer, Heidelberg, Germany, Nov.~16--19,
  2020, pp. 261--290.

\bibitem{CCS:ACDT21}
J.~Alwen, S.~Coretti, Y.~Dodis, and Y.~Tselekounis, ``Modular design of secure
  group messaging protocols and the security of {MLS},'' in \emph{ACM CCS 2021:
  28th Conference on Computer and Communications Security}, G.~Vigna and
  E.~Shi, Eds.\hskip 1em plus 0.5em minus 0.4em\relax Virtual Event, Republic
  of Korea: {ACM} Press, Nov.~15--19, 2021, pp. 1463--1483.

\bibitem{whatsapp}
WhatsApp, ``{WhatsApp Encryption Overview Technical white paper, v.3},'' Oct.
  2020,
  \url{https://www.whatsapp.com/security/WhatsApp-Security-Whitepaper.pdf}.

\bibitem{signalrepo}
\BIBentryALTinterwordspacing
M.~Marlinspike \emph{et~al.}, ``Signal protocol.'' [Online]. Available:
  \url{https://github.com/signalapp/libsignal-protocol-java/tree/master/java/src/main/java/org/whispersystems/libsignal}
\BIBentrySTDinterwordspacing

\bibitem{CSF:CohCreGar16}
K.~{Cohn-Gordon}, C.~J.~F. Cremers, and L.~Garratt, ``On post-compromise
  security,'' in \emph{CSF 2016: IEEE 29st Computer Security Foundations
  Symposium}, M.~Hicks and B.~Köpf, Eds.\hskip 1em plus 0.5em minus
  0.4em\relax Lisbon, Portugal: {IEEE} Computer Society Press, jun~27-1 2016,
  pp. 164--178.

\bibitem{admins}
D.~Balb{á}s, D.~Collins, and S.~Vaudenay, ``{Cryptographic Administrators for
  Secure Group Messaging},'' Cryptology ePrint Archive.

\bibitem{USENIX:CreHalKoh21}
C.~Cremers, B.~Hale, and K.~Kohbrok, ``The complexities of healing in secure
  group messaging: Why cross-group effects matter,'' in \emph{USENIX Security
  2021: 30th {USENIX} Security Symposium}, M.~Bailey and R.~Greenstadt,
  Eds.\hskip 1em plus 0.5em minus 0.4em\relax {USENIX} Association,
  Aug.~11--13, 2021, pp. 1847--1864.

\bibitem{x3dh}
\BIBentryALTinterwordspacing
M.~Marlinspike and T.~Perrin, ``The x3dh key agreement protocol,'' 2016.
  [Online]. Available:
  \url{https://signal.org/docs/specifications/x3dh/x3dh.pdf}
\BIBentrySTDinterwordspacing

\bibitem{KraBelCan97}
H.~Krawczyk, M.~Bellare, and R.~Canetti, ``{HMAC}: Keyed-hashing for message
  authentication,'' IETF Internet Request for Comments 2104, Feb. 1997.

\bibitem{RFC5869}
\BIBentryALTinterwordspacing
H.~Krawczyk and P.~Eronen, ``Hmac-based extract-and-expand key derivation
  function (hkdf),'' Internet Requests for Comments, RFC Editor, RFC 5869, May
  2010, \url{http://www.rfc-editor.org/rfc/rfc5869.txt}. [Online]. Available:
  \url{http://www.rfc-editor.org/rfc/rfc5869.txt}
\BIBentrySTDinterwordspacing

\bibitem{rosler2018more}
P.~Rösler, C.~Mainka, and J.~Schwenk, ``More is less: On the end-to-end
  security of group chats in signal, whatsapp, and threema,'' in \emph{2018
  IEEE European Symposium on Security and Privacy (EuroS\&P)}, 2018, pp.
  415--429.

\bibitem{PROVSEC:LaMLauMit07}
B.~A. LaMacchia, K.~Lauter, and A.~Mityagin, ``Stronger security of
  authenticated key exchange,'' in \emph{ProvSec 2007: 1st International
  Conference on Provable Security}, ser. Lecture Notes in Computer Science,
  W.~Susilo, J.~K. Liu, and Y.~Mu, Eds., vol. 4784.\hskip 1em plus 0.5em minus
  0.4em\relax Wollongong, Australia: Springer, Heidelberg, Germany, Nov.~1--2,
  2007, pp. 1--16.

\end{thebibliography}
